\documentclass[prx,amssymb,superscriptaddress,twocolumn,longbibliography]{revtex4-2}
\pdfoutput=1
\usepackage[most]{tcolorbox} 
\usepackage{xcolor} 
\usepackage[utf8]{inputenc}
\usepackage[english]{babel}
\usepackage{tikz}
\usetikzlibrary{shapes,backgrounds,fit,decorations.pathreplacing,arrows,decorations.markings}
\usepackage{verbatim}
\tikzstyle{vecArrow} = [thick, decoration={markings,mark=at position
   1 with {\arrow[semithick]{open triangle 60}}},
   double distance=1.4pt, shorten >= 5.5pt,
   preaction = {decorate},
   postaction = {draw,line width=1.4pt, white,shorten >= 4.5pt}]
\tikzstyle{innerWhite} = [semithick, white,line width=1.4pt, shorten >= 4.5pt]
\usepackage{bbm}
\usepackage{bm}
\usepackage{amsbsy}
\usepackage{amsthm}
\usepackage{amssymb}
\usepackage{amsfonts}
\usepackage{amsmath}
\usepackage{dsfont} 
\usepackage{graphicx} 
\usepackage{epsfig}
\usepackage{epstopdf}
\usepackage{dsfont}
\usepackage{booktabs,multirow} 
\usepackage{color}

\usepackage[colorlinks]{hyperref}
\makeatletter
\newcommand\org@hypertarget{}
\let\org@hypertarget\hypertarget
\renewcommand\hypertarget[2]{%
  \Hy@raisedlink{\org@hypertarget{#1}{}}#2%
  }
\makeatother
\usepackage[figure,table]{hypcap}
\usepackage{MnSymbol}
\usepackage{enumerate}
\usepackage
{enumitem}
\usepackage{float}
\hypersetup{
	bookmarksnumbered,
	pdfstartview={FitH},
	citecolor={darkgreen},
	linkcolor={darkred},
	urlcolor={darkblue},
	pdfpagemode={UseOutlines}}
\definecolor{darkgreen}{RGB}{50,190,50}
\definecolor{darkblue}{RGB}{0,0,190}
\definecolor{darkred}{RGB}{238,0,0}
\definecolor{quantum}{RGB}{83,37,127}
\definecolor{quantumlight}{RGB}{169,146,191}
\usepackage{soul}

\newcommand{\ket}[1]{\ensuremath{\left|\right.\!{#1}\!\left.\right\rangle}}

\newcommand{\ketbra}[2]{\ensuremath{|{#1}\rangle\!\langle{#2}|}}

\renewcommand{\thesection}{\Roman{section}}
\renewcommand{\thesubsection}{\Roman{section}.
\arabic{subsection}}

\begin{document}

\title{Quantum detectors as autonomous machines: assessing the nonequilibrium thermodynamics of information acquisition}
\author{Emanuel Schwarzhans}
\email{emanuel.schwarzhans@um.edu.mt}
\affiliation{Department of Physics, University of Malta, Msida MSD 2080, Malta }
\affiliation{Atominstitut, TU Wien, 1020 Vienna, Austria}
\author{Tony~J.~G.~Apollaro}
\email{tony.apollaro@um.edu.mt}
\affiliation{Department of Physics, University of Malta, Msida MSD 2080, Malta }
\author{Ilia Khomchenko}
\email{ilia.khomchenko@um.edu.mt}
\affiliation{Department of Physics, University of Malta, Msida MSD 2080, Malta }
\author{Maximilian~P.~E.~Lock}
\email{maximilian.paul.lock@tuwien.ac.at}
\affiliation{Atominstitut, TU Wien, 1020 Vienna, Austria}
\affiliation{Institute for Quantum Optics and Quantum Information (IQOQI),
\\Austrian Academy of Sciences, 1090 Vienna, Austria}

\author{Mark T. Mitchison}
\email{mark.mitchison@kcl.ac.uk}
\affiliation{School of Physics, Trinity College Dublin, College Green, Dublin 2, D02 K8N4, Ireland}
\affiliation{Department of Physics, King’s College London, Strand, London, WC2R 2LS, United Kingdom}

\author{Marcus Huber}
\email{marcus.huber@tuwien.ac.at}
\affiliation{Atominstitut, TU Wien, 1020 Vienna, Austria}
\affiliation{Institute for Quantum Optics and Quantum Information (IQOQI),
\\Austrian Academy of Sciences, 1090 Vienna, Austria}

\begin{abstract}
We formulate a minimal model of a quantum particle detector as an autonomous quantum thermal machine. Our goal is to establish how entropy production, which is needed to maintain the detector out of equilibrium, is linked to the quality of the measurement process. Using our model, we perform a detailed investigation of the detector's key performance characteristics: namely, detection efficiency, gain, jitter, dead time, and dark counts. We find that entropy production constrains both the efficiency and temporal precision of the detection process, in the sense that improved performance generally requires more dissipation. We also find that reducing either the detection jitter or dead time unavoidably increases the rate of dark counts. Our work establishes a quantitative connection between entropy production and the quality of the irreversible detection process, highlights fundamental tradeoffs in the performance of particle detectors, and provides a framework for further investigations of the non-equilibrium thermodynamics of quantum measurement and amplification. 
\end{abstract}


\maketitle

\section{Introduction}

\begin{figure*}[t]
    \centering
    \includegraphics[width=\textwidth]{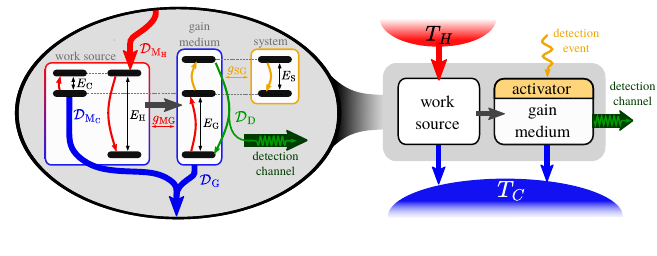}
    \caption{Illustration of the minimal model for single event detection. The right side depicts a rough overview of the necessary elements for autonomous single-event amplification, \emph{i.e.}~a \emph{gain medium} and a \emph{work source}, and a quantum system to be detected (indicated by the vertical yellow wavy arrow). The left side depicts the detailed energy level structure constituting the minimal model that achieves this task, and all parameters that determine a detector's performance, \emph{i.e.}~the energy gaps $E_\mathrm{C}$, $E_H$ of the machine's cold and hot qubit, $E_\mathrm{G}=E_H-E_\mathrm{C}$ of the gain medium's first gap and the energy $E_\mathrm{S}$ of the system to be detected (yellow box), which is resonant with the gain mediums second gap, the machine-gain medium coupling $g_\mathrm{MG}$, the gain medium-system coupling $g_\mathrm{SG}$ and the dissipator coupling to the respective baths and the detector channel $\mathcal{D}_{M_H}$, $\mathcal{D}_{M_C}$, $\mathcal{D}_\mathrm{G}$ and $\mathcal{D}_\mathrm{D}$. }
    \label{fig:overview_illustration}
\end{figure*}

The process of measurement is an essential step in any form of information processing. In quantum systems, measurement requires information about a microscopic property to be reproduced by a macroscopic detector so that it can be recorded or broadcast for further use. And, as with all information processing, measurement comes with thermodynamic costs and limitations. An early example was Landauer's discovery that erasing one bit of information necessarily entails a minimum amount of heat dissipation~\cite{landauer1961}. While this idealised bound is not reachable for practical finite-time processes \cite{Proesmans2020,Miller2020,Zhen2021,Vu2022, taranto2023b, Rolandi2023}, it nonetheless implies a fundamental work cost for any quantum measurement, which must be invested to either initially prepare~\cite{guryanova2020} or ultimately erase~\cite{faist2015a,Chu2022} the memory that stores the outcome. In recent years, other thermodynamic limitations of quantum measurement have been studied intensively, including the work cost of correlating the system and the memory~\cite{Sagawa2009,Granger2011,Jacobs2012, Latune2025}, the importance of a phase transition to amplify the measurement outcome~\cite{allahverdyan_quantum_2001, allahverdyan_curie-weiss_2003}, and the irreversible nature of ``objectification''~\cite{schwarzhans2023a,engineer2024}: the process whereby information becomes redundantly encoded across many classical degrees of freedom~\cite{zurek_quantum_2014, Horodecki2015}. 

The vast majority of prior investigations into the thermodynamics of measurement have taken place against the backdrop of a single thermal environment. In this context, any energetic or entropic costs are associated with the initial departure from, and eventual return to, thermal equilibrium by the system or measurement apparatus. This paradigm fails to capture the basic working principles of many real detectors, whose operation requires them to be maintained in a nonequilibrium steady state (NESS). Whether considering charge detection by a quantum point contact~\cite{Vigneau2023}, microwave-field detection using a quantum-limited linear amplifier~\cite{Blais2021}, or photodetection by an avalanche photodiode~\cite{Shi2024}, any continuous quantum measurement with high gain demands a steady flux of energy through the detector. Indeed, it is known that a quantum-limited detector in the linear-response regime exhibits strong non-reciprocity between its input and output~\cite{Clerk2003}, thus violating the Onsager reciprocity conditions and implying that a good detector must be far from equilibrium~\cite{clerk_introduction_2010}. This suggests an intriguing connection between entropy production within the detector and the quality of the measurement process. However, the general relationship between heat dissipation and detector performance remains largely unexplored.

Here, we address this problem
by constructing a minimal model of a detector that is fully self-contained and autonomous, allowing us to capture all thermodynamic resources necessary for its operation. A similar strategy has proven fruitful in establishing fundamental limits on timekeeping~\cite{erker2017b,schwarzhans_autonomous_2021} and refrigeration~\cite{Linden2010, Levy2012}. So let us set out by defining the minimal properties that any such detector model must exhibit:
\begin{enumerate}[label=P.\arabic*.]
    \item It must be based on a physical interaction between a target and a measurement device \cite{wiseman2009,vonneumann1932}.
    \label{enum:physical_interaction}
    \item It must be autonomous, \emph{i.e.}~without external (temporal) control to mediate the measurement.
    \label{enum:autonomous}
    \item It should be usable repeatedly or continuously. \label{enum:repeatability}
    \item It needs to \emph{amplify} the signal for a macroscopically noticeable state change to occur.\label{enum:amplify}
\end{enumerate}
We focus on detectors that produce a discrete, macroscopic ``click'' when a quantum particle is present, e.g.~a photodetector, as opposed to those that produce a continuous signal, e.g.~a quantum-limited field amplifier or heterodyne detector. To characterise the performance of such particle detectors, we identify four desiderata that an ``ideal" detector would have to fulfil \cite{hadfield2009}:
\begin{tcolorbox}[colback=purple!5!white, colframe=purple!75!black, title=\textbf{Desiderata}]
\begin{enumerate}[label=D.\arabic*.]
    \item It clicks \emph{only} when a quantum excitation is present; clicks from fluctuations are called \mbox{\emph{dark counts}}, and are given by the dark count rate $R_\mathrm{dc}$. \label{enum:dark_counts}
    \item These clicks should happen at the \emph{exact time} a quantum excitation is present; uncertainty in this timing is called \mbox{\emph{detection jitter}} $\Delta t_\mathrm{D}$.\label{enum:jitter} 
    \item It should \emph{always} click when a quantum excitation is present; the actual success rate is called \mbox{\emph{detection efficiency}} $\eta_\mathrm{D}$. \label{enum:efficiency}
    \item After each detection, it should be \emph{immediately} ready for the next; the minimal recovery time is called \emph{dead time} $D$.
    \label{enum:deadtime}
\end{enumerate}
\end{tcolorbox}

In Sec.~\ref{sec:model}, we outline a minimal model that exhibits the properties~\ref{enum:physical_interaction}--\ref{enum:amplify}. As illustrated in Fig.~\ref{fig:overview_illustration}, our setup exploits a quantum thermal absorption machine~\cite{Mitchison2019} to drive a gain medium into a metastable state, which is triggered by an incoming particle to produce an amplified output current with high energy gain. We then define metrics that quantify how well each of the desiderata~\ref{enum:dark_counts}--\ref{enum:deadtime} is satisfied in Sec.~\ref{sec:performance_metrics}, before applying them to analyse the performance of our minimal detector model in Sec.~\ref{sec:results}. We demonstrate that higher detection efficiency comes at the price of increased entropy production. We also find a tradeoff between the detector's dark count rate and detection jitter, where the latter can be decreased at the expense of the former by increased heat dissipation. These results establish a quantitative link between the irreversible nature of the detection process and the dissipation of heat in the non-equilibrium steady state of the detector. We discuss the consequences of these findings for the foundations of quantum thermodynamics and quantum measurement theory in Sec.~\ref{sec:conclusions}.

\section{The Model}
\label{sec:model}

Real-world detectors typically consist of many-body systems comprising multiple energy levels that contribute to generating an amplified signal with macroscopically detectable energy. To achieve this, independent of their explicit physical implementation, all detectors share a common feature—a metastable high-energy initial state.

To facilitate a tractable analysis of fundamental thermodynamic limitations, we abstract these many-body dynamics into a minimal effective model that requires no external control to operate (no active on/off switching). The minimal model must consist of two ingredients: a \emph{work source}, modelled as a two-qubit quantum thermal machine operating in the heat engine regime  (\emph{i.e.}, inducing a population inversion) \cite{brunner2012}, utilising a heat flow from a hot to a cold bath to provide work to \emph{gain medium}, which is modelled as a three-level system. The \emph{work source} drives the \emph{gain medium} out of equilibrium into a high-energy state, which functions as a ``metastable" energy level (see Fig.~\ref{fig:overview_illustration}). The task of the gain medium is twofold: it facilitates interaction with the quantum particle that triggers the detection event (\emph{e.g.}, the photon), and it amplifies the signal produced by such a detection to become macroscopically detectable. 

Reducing the size of the model further is not feasible, since amplification requires the gain medium to feature at least three distinct energetic states: The ``ground state" $\ket{0}_\mathrm{G}$, the ``metastable" (or ``detection-ready") state $\ket{1}_\mathrm{G}$ which is at a high energy compared to the energy of the quantum particle, and the ``activated" state $\ket{2}_\mathrm{G}$ which can only be accessed via interaction with a quantum particle at the right energy.

Ideally, a detection event lifts the gain medium from the ``detection-ready" state into the ``activated" state after which it decays via the detection channel into the ground state. The energy dissipated in this manner corresponds to the \emph{threshold energy} and is assumed to be macroscopically detectable (or sufficiently large to trigger a von Neumann chain \cite{vonneumann1932}). Once the gain medium reaches the ground state, the work source lifts it back into the detection-ready state, in which it remains until the next quantum particle arrives, as depicted in Fig.~\ref{fig:overview_illustration}. However, since every part of the detector is necessarily embedded in a thermal environment, thermal fluctuations introduce noise, reducing the detector's performance.

We model the quantum particle to be detected as a two-level system that interacts coherently with the gain medium, where $\ket{0}_\mathrm{S}$ means no particle is present and $\ket{1}_\mathrm{S}$ means there is a particle. 

The total Hilbert space structure of our model is given as $\mathcal{H}=\mathcal{H}_\mathrm{M_C}\otimes\mathcal{H}_\mathrm{M_H}\otimes\mathcal{H}_\mathrm{G}\otimes\mathcal{H}_S$, where $\mathcal{H}_\mathrm{M_C}$ and $\mathcal{H}_\mathrm{M_H}$ denote the space of the hot and the cold qubit (jointly the machine subspace $\mathcal{H}_\mathrm{M}=\mathcal{H}_\mathrm{M_C}\otimes\mathcal{H}_\mathrm{M_H}$), $\mathcal{H}_\mathrm{G}$ the gain medium and $\mathcal{H}_S$ denotes the space of the quantum particle (e.g. that of a photon).

The free Hamiltonian of the total model is given as
\begin{align}
    &H_\mathrm{M}=E_\mathrm{C}\ketbra{1}{1}_\mathrm{M_C} +(E_\mathrm{C}+E_\mathrm{G})\ketbra{1}{1}_\mathrm{M_H}
    \\&
    H_\mathrm{G}= E_\mathrm{G} \ketbra{1}{1}_\mathrm{G}+ (E_\mathrm{G} +E_\mathrm{S})\ketbra{2}{2}_\mathrm{G}
    \\&
    H_\mathrm{S}=E_\mathrm{S} \ketbra{1}{1}_\mathrm{S}
    \\&
    H_0=H_\mathrm{M}+H_\mathrm{G}+H_\mathrm{S},
\end{align}
The energy level structure of each of the model's components is illustrated in Fig.~\ref {fig:overview_illustration}.

To ensure a complete bookkeeping of all thermodynamic resources, we require energy-conserving interactions, \emph{i.e.}~that $[H,H_0]=0$ (this also avoids the additional noise otherwise mandated by the Wigner–Araki–Yanase theorem for symmetry-breaking observables~\cite{ozawa2002conservation}). Therefore, the transition $\ket{0}_\mathrm{G}$ to $\ket{1}_\mathrm{G}$, driven by the quantum thermal machine, must be resonant with the virtual qubit of the thermal machine. The transition from $\ket{1}_\mathrm{G}$ to $\ket{2}_\mathrm{G}$ needs to be resonant with the energy of the quantum particle. This leaves us with the total interaction Hamiltonian 
\begin{align}
    H_I=g_\mathrm{MG}\ketbra{101}{010}_\mathrm{M_C, M_H, G} +  g_\mathrm{SG}\ketbra{20}{11}_\mathrm{G, S} + h.c.
\end{align}
The total Hamiltonian is $H=H_0+H_I$. The dynamics of the total system coupling to a hot and a cold bath, as indicated by Fig.~\ref{fig:overview_illustration}, is given by the Lindblad master equation
\begin{align}\label{eq:Lindbladian}
    \mathcal{L}\rho=-i[H,\rho] + (\mathcal{D}_\mathrm{M_C} +\mathcal{D}_\mathrm{M_H}+\mathcal{D}_\mathrm{G}+\mathcal{D}_\mathrm{D})\rho,
\end{align}
where the dissipators and the corresponding dissipation rates are given explicitly in App.~\ref{appsec:Lindbladian}. $\mathcal{D}_\mathrm{M_{C/H}}$ are Lindblad dissipators connecting the hot and cold machine qubits with their respective baths at temperatures $T_C$ and $T_H$, $\mathcal{D}_\mathrm{G}$ acts to thermalise the gain medium with the environment, which we assume to be at the cold bath temperature $T_C$, and the detection channel is implemented via the dissipator
\begin{align}\mathcal{D}_\mathrm{D}=\Gamma^+_\mathrm{D}(T_\mathrm{C}) \, \mathcal{D}[L_\mathrm{D}^+]+\Gamma^-_\mathrm{D}(T_\mathrm{C})\, \mathcal{D}[L_\mathrm{D}^-]~.
\end{align} 
Here, $\mathcal{D}[L]\rho=L \rho L^{\dagger}-\frac{1}{2}\left\{L^{\dagger} L, \rho\right\}$ with $L_\mathrm{D}^+=\ketbra{2}{0}_\mathrm{G}$ and  $L_\mathrm{D}^-=\ketbra{0}{2}_\mathrm{G}$, while $\Gamma_\mathrm{D}^\pm(T_\mathrm{C})$ are the temperature-dependent gain and loss rates. This is the channel which provides information about the detection; it couples the lowest level of the gain medium to its maximally excited energy level, \emph{i.e.}~$\ket{0}_\mathrm{G}$ and $\ket{2}_\mathrm{G}$.

\section{Transient behaviour and detection performance}
\label{sec:performance_metrics}
Under the dynamics given by Eq.~\eqref{eq:Lindbladian}, the detector will evolve towards the unique non-equilibrium steady state  $\rho^{ss}$, defined by $\mathcal{L}\rho^{ss}=0$. This can be understood as the detection-ready state.

Note that the dynamics do not include incoming particles. To trigger a detection event, we thus initialise the detector in the state $\rho^{ss}$ and then bring in a particle, corresponding to the update
\begin{align}\label{eq:init}
    \rho_0=\dfrac{\sigma^+_\mathrm{S} \rho^{ss}\sigma^-_\mathrm{S}}{ \operatorname{Tr(\sigma^-_\mathrm{S}\sigma^+_\mathrm{S}\rho^{ss})}}=\operatorname{Tr}_\mathrm{S}[\rho^{ss}]\otimes \ketbra{1}{1}_\mathrm{S},
\end{align}
\emph{i.e.}~replacing S with an excited state. We can see this as conditioning on the case where a particle is present (see~\cite{landi2024}, section II.C.). We then let the total system evolve via Eq.~\eqref{eq:Lindbladian} until the detector has returned to the detection-ready state $\rho^{ss}$.

The detector's performance, quantified by how well it achieves the desiderata \ref{enum:dark_counts}-\ref{enum:deadtime}, is determined by the transient behaviour of the average net number of excitations emitted via the detection channel $\mathcal{D}_\mathrm{D}$, \emph{i.e.}~the average \emph{detection current} 
\begin{align}
    J_\mathrm{D}(t)=\operatorname{Tr}\left[\mathcal{J}_\mathrm{D} \rho(t)\right]=\operatorname{Tr}\left[( \Gamma_\mathrm{D}^- \ketbra{2}{2}_\mathrm{G}- \Gamma_\mathrm{D}^+ \ketbra{0}{0}_\mathrm{G})\rho(t) \right].
\end{align}
The first term above corresponds to the emission of an amplified excitation by the detector, and thus a successfully registered detection event. The second term describes unwanted absorption by the detector from the thermal background. $\mathcal{J}_\mathrm{D}$ is the superoperator describing quantum jumps in the detection channel~\cite{landi2024}, defined as
\begin{align}
    \mathcal{J}_\mathrm{D}\rho= \Gamma_\mathrm{D}^-(T_\mathrm{C})\,L_\mathrm{D}^-\rho L_\mathrm{D}^+ -\Gamma_\mathrm{D}^+(T_\mathrm{C})\,L_\mathrm{D}^+\rho L_\mathrm{D}^-,
\end{align}
and $\Gamma_\mathrm{D}^\pm(T_\mathrm{C})$ are the temperature-dependent coupling rates (see App.~\ref{appsec:Lindbladian}).
The emitted excitations are greater in energy than that of the incoming particle which triggered them. The magnitude of this amplification is given by the \emph{gain} $G=\tfrac{E_\mathrm{G}}{E_\mathrm{S}}$ (see App.~\ref{appsec:gain}).

We treat consecutive detection events as independent; in particular we assume that the detector reaches $\rho^{ss}$ again before the next particle arrives and that there are no memory effects between individual runs. Thus, the average of a single event is equivalent to averaging over many consecutive detection events, allowing us to consider only one run.

\begin{figure}
    \centering
    \includegraphics[width=1\linewidth]{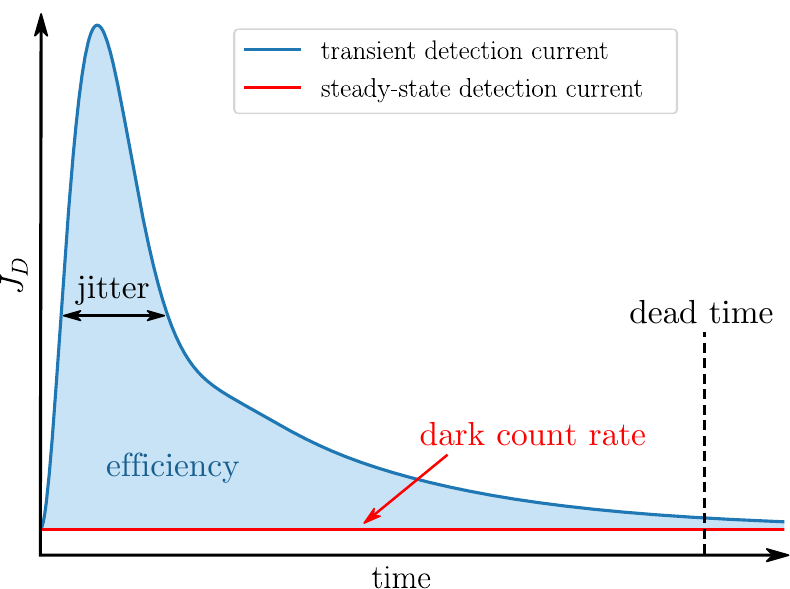}
    \caption{Schematic dynamics of one detection event. The blue line shows the transient current in the detection channel $\mathcal{D}_\mathrm{D}$. The red line indicates the dark current given by the steady-state current $J_\mathrm{D}^{ss}$. Integrating the excess detection current gives the detection efficiency $\eta_\mathrm{D}$, as indicated by the blue area. The detection jitter $\Delta t_\mathrm{D}$ is given by the variance of the excess detection current, as indicated by the horizontal arrows in the blue area. The dead time $D$ is the timescale of returning to the detection-ready state $\rho^{ss}$, indicated by the dotted vertical line. 
    }
    \label{fig:dynamics}
\end{figure}

We can now formalise the figures of merit corresponding to the desiderata. Fig.~\ref{fig:dynamics} provides a schematic example of the detection current for one detection event and the corresponding figures of merit.  

The \emph{dark count rate} (\ref{enum:dark_counts}) is given as the steady-state detection current of the detector, representing the false-positive detection rate
\begin{align}
    &     R_\mathrm{dc}=J^{ss}_\mathrm{D}=\operatorname{Tr}\left[\mathcal{J}_\mathrm{D} \rho^{ss}\right].
\end{align}
The \emph{detection jitter} (\ref{enum:jitter}) is given as the temporal variance of the normalised excess detection current $p_{\tilde{J}_\mathrm{D}}(t)=\tfrac{1}{\eta_\mathrm{D}}(J_\mathrm{D}(t)-J^{ss}_\mathrm{D})$:
\begin{align}
    &  \Delta t_\mathrm{D}=\operatorname{Var}(t)_{p_{\tilde J_\mathrm{D}}} =\int_0^\infty dt\,t^2  \frac{\tilde J_\mathrm{D}(t)}{\eta_\mathrm{D}} -\left[\int_0^\infty dt\,t \frac{\tilde J_\mathrm{D}(t)}{\eta_\mathrm{D}}\right]^2
\end{align}
where $\tilde{J}_D(t)=J_D(t)-J_D^{ss}$  and $\eta_\mathrm{D}$ is the \emph{detection efficiency} (\ref{enum:efficiency}), given as the total (excess) number of jumps through the detection channel
\begin{align}
    & \eta_\mathrm{D}=\int_0^\infty dt\tilde{J}_\mathrm{D}(t).
\end{align}

Finally, the \emph{dead time} (\ref{enum:deadtime}) is defined as the timescale at which the detector returns to the steady state (\emph{i.e.}~the time it takes to reset), which we take to be the slowest timescale in the dynamics, given by the smallest non-zero eigenvalue of the Liouvillian, denoted $\lambda_1$:
\begin{align}\label{eq:epsilon_dead_time} 
    D\sim -\dfrac{1}{\operatorname{Re}(\lambda_1)}
\end{align}
note that $\operatorname{Re}(\lambda_1)< 0$.

For further details on the figures of merit, please refer to App.~\ref{appsec:figuresofmerit}.

\section{Performance trade-off relations}
\label{sec:results}

Any detector requires some resources to operate. As mentioned, our detector acquires energy from a work source, which it inevitably loses, either due to the necessary coupling to the environment or due to its conversion into a macroscopically detectable signal. The environmental coupling is responsible for these energy losses or dissipation, which inevitably result in entropy production. 
Increasing the precision of the output of a stochastic process typically comes at the cost of increased entropy production \cite{barato2015a,gingrich2016,hasegawa2020}. Thus, one would generally expect the performance of detectors, and thus their capacity to fulfil the desiderata \ref{enum:dark_counts}-\ref{enum:deadtime} to come at an entropic cost. Additionally, optimising one figure of merit can, in principle, come at the cost of worsening others, resulting in trade-off relations between some of these quantities. 

In the following, we will present such trade-offs revealed by numerical sampling of the entire parameter space (see Fig.~\ref{fig:overview_illustration}), for a given fixed energy gain $G=9$ (see App.~\ref{appsec:gain} for scaling behaviour with $G$). This, in turn, allows us to make informed proposals for analytic relations (see \cite{Schwarzhans_Quantum_detectors_as} for all datasets and the complete code used for simulation).

The missing ingredient for understanding thermodynamic costs related to measurements is the total entropy dissipated during a single detection event $\Sigma_\mathrm{tot}$. It can be split into two parts: the entropy production rate required to keep the detector ready, 
\begin{align}
    \dot\Sigma_{ss}=\!\dfrac{1}{T_\mathrm{C}}\operatorname{Tr}\left[ (E_\mathrm{C} \mathcal{J}_\mathrm{M_C}+ E_\mathrm{G}  \mathcal{J}_\mathrm{G_{1-0}}+E_\mathrm{S}\mathcal{J}_\mathrm{G_{2-1}})\rho^{ss}\right],
\end{align}
where $\mathcal{J}_\mathrm{M_C}$, $\mathcal{J}_{L_{1-0}}$ and $\mathcal{J}_{L_{2-1}}$ denote the current operators related to dissipation into the cold bath,
and the additional entropy produced by the transient dynamics
\begin{align}
    \Sigma_\mathrm{trans}\!=\!\dfrac{1}{T_\mathrm{C}}\!\int_0^\infty\!\!\!\!\!\! dt \left( E_\mathrm{C} \tilde J_\mathrm{M_C}(t) + E_\mathrm{G} \tilde J_\mathrm{G_{1-0}}(t)+E_\mathrm{S}\tilde J_\mathrm{G_{2-1}}(t)\right),
\end{align}
where $\tilde J_x(t) =J_x(t)-J_x^{ss}$ is the excess current thorugh channel $\mathcal{D}_x$ with index $\mathrm{G_{1-0}}$ and $\mathrm{G_{2-1}}$ corresponding to the channels of the first and second gap of G.
Accordingly, we write:
\begin{align}
    \Sigma_\mathrm{tot}\sim -\frac{\dot\Sigma_{ss}}{\operatorname{Re}(\lambda_1)} + \Sigma_\mathrm{trans},
\end{align}
where the entropy production due to the steady-state entropy production rate is calculated as $\int_0^D dt\dot\Sigma_{ss}\sim-\frac{\dot\Sigma_{ss}}{\operatorname{Re}(\lambda_1)}$ (see eq.~\eqref{eq:epsilon_dead_time}). For further details, see App.~\ref{appsec:figuresofmerit}.

Now we can ask: \emph{Does the entropy production impose a constraint on the efficiency $\eta_\mathrm{D}$?}
\begin{figure}
    \centering
    \includegraphics[width=1\linewidth]{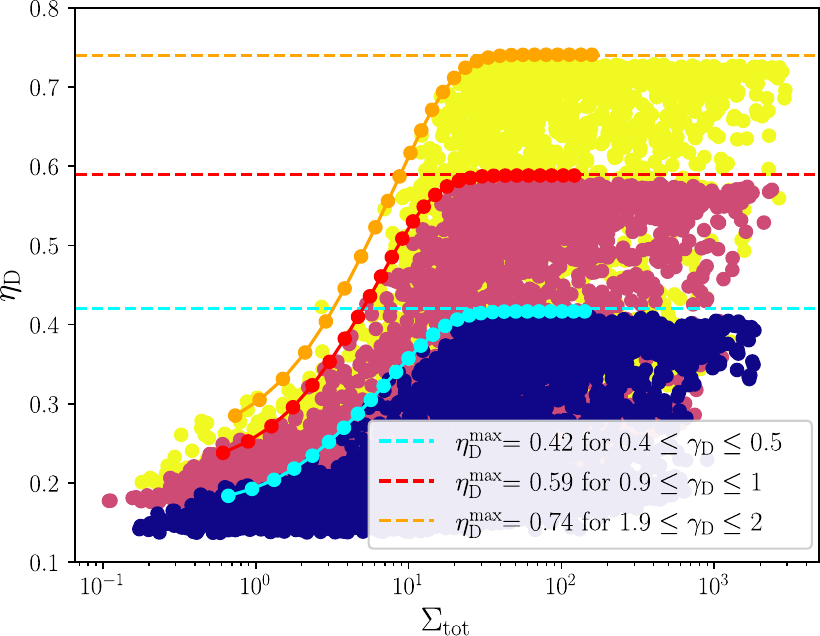}
    \caption{Detection efficiency $\eta_\mathrm{D}$ over total entropy production $\Sigma_\mathrm{tot}$ for different values of detection channel coupling $\gamma_\mathrm{D}\in \{0.45,0.95,1.95\}$ (selecting points in a range $\pm0.05$ around these values for the scatter plots) and fixed bath coupling $\gamma_\mathrm{G}=0.7$. The horizontal dashed lines indicate the maximal efficiency for the given coupling rates given by Eq.~\eqref{eq:max_efficiency}. The lines with dots are along $T_\mathrm{V}=\tfrac{-T_\mathrm{C} E_H}{E_\mathrm{C}}(1+\epsilon)$, where $\epsilon=10^{-6}$, i.e. close to the maximal virtual temperature for which the thermal machine still operates in the engine regime\cite{brunner2012}. The parameter ranges for the scatter plots are: $T_\mathrm{C}\in [0.05,1]$, $T_\mathrm{V}\in[-10,\tfrac{-T_\mathrm{C} E_H}{E_\mathrm{C}}]$, $g_\mathrm{SG}\in[0.1,1]$, $g_\mathrm{MG}\in[0.1,2]$, $\gamma_\mathrm{M}\in[0.1,2]$ and $f_{E_\mathrm{C}}\in [0.1,2]$ (where $E_\mathrm{C}=f_{E_\mathrm{C}}E_\mathrm{G}$). }
    \label{fig:Efficiency_entropy_production_rateD}
\end{figure}
Within our model, this question can be answered positively. To achieve a certain efficiency, the detector must necessarily dissipate a minimal amount of entropy. The quantitative behaviour is depicted in Fig.~\ref{fig:Efficiency_entropy_production_rateD}, via scatter plots over all parameters (see Fig.~\ref{fig:overview_illustration}), except for $\gamma_\mathrm{D}$ and $\gamma_\mathrm{G}$, whose ratio is kept constant. We see that there is a regime in which increased efficiency requires an increase in entropy production per detection event.
Increasing the entropy production beyond that will yield reduced efficiency gains, eventually resulting in a plateau. In App.~\ref{appsec:Efficiency_bound} we conjecture that the maximal efficiency is given by:
\begin{align}\label{eq:max_efficiency}
    \eta_\mathrm{max}=\dfrac{1}{1+ \frac{\gamma_\mathrm{G}}{\gamma_\mathrm{D}}},
\end{align}
which is achieved in the limit $T_\mathrm{C}\rightarrow 0$, \emph{i.e.}~diverging entropy production.
This is supported by the numerical results, which can be found in Fig.~\ref{fig:Efficiency_entropy_production_rateD}. The rationale behind this hypothesis is the following:
In the conditional branch where a particle is present, \emph{i.e.}~where the detector is in a state given by eq.~\eqref{eq:init}, the excitation from the system can coherently be transferred between the system and the gain medium, or dissipatively leave the gain medium-system subspace spanned by $\mathds{1}_\mathrm{M}\ket{1}_\mathrm{G}\otimes\ket{1}_S$ and $\mathds{1}_\mathrm{M}\ket{2}_\mathrm{G}\otimes\ket{0}_S$. Thus, the excitation can only leave via the detection channel $\mathcal{D}_\mathrm{D}$ or the thermalising channel $\mathcal{D}_\mathrm{G}$, which leads us to the guess that the coupling rates of these channels determine the efficiency.
We observe numerically that this heuristic does not hold in the finite temperature case, but it does in the limit $T_\mathrm{C}\rightarrow 0$.
Since $\gamma_\mathrm{G}$ is small if the detector is well isolated from its environment and $\gamma_\mathrm{D}$ depends on our ability to couple it to the monitored environment, the maximally achievable detection efficiency is limited by our ability to isolate the gain medium from the environment, even if the environment is at zero temperature. 

\begin{figure}
    \centering
    \includegraphics[width=1\linewidth]{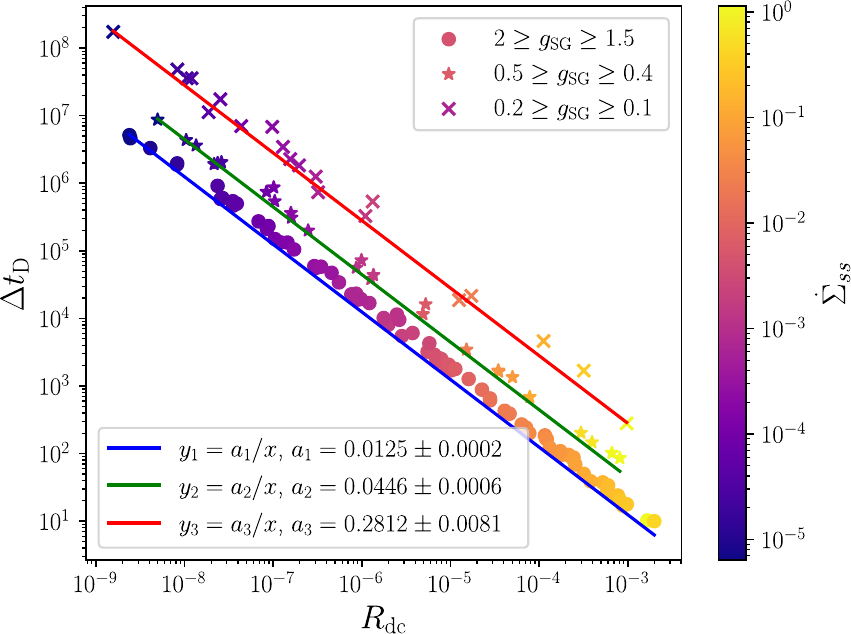}
    \caption{ Detection jitter $\Delta t_\mathrm{D}$ over dark count rate $R_{dc}$
    close to the maximum efficiency $0.72<\eta_\mathrm{D}<\eta_\mathrm{D}^\mathrm{max}\approx0,74$. The colour map shows the steady-state entropy production rate. The solid lines are fits through these data points of $y=a/x$ where $y=R_{dc}$ and $x=-1/\operatorname{Re}(\lambda_1)$. The parameter range is the same as in Fig.~\ref{fig:Efficiency_entropy_production_rateD}, except for $\gamma_\mathrm{D}\in[0.1,2]$ and $g_\mathrm{SG}\in\{[0.1,0.2],[0.4,0.5],[1.5,2]\}$. }

    \label{fig:jitter_over_dark_count_rate_2}
\end{figure}

High detection efficiency alone, however, is not sufficient for most information processing tasks. These often require precise temporal information (\emph{i.e.}~low \emph{detection jitter} $\Delta t_\mathrm{D}$) and a low rate of counting false positives (\emph{i.e.}~low \emph{dark count rate} $R_\mathrm{dc}$). Given a highly efficient detector, we find that jitter and dark count rate can, in fact, not be improved independently -- they follow a trade-off relation of the form (see Fig.~\ref{fig:jitter_over_dark_count_rate_2})
\begin{align}\label{eq:jitter-darkcounts}
    \Delta t_\mathrm{D} \propto \dfrac{1}{R_\mathrm{dc}}
\end{align} 
Although it is well known that the dark count rate can be mitigated through gating, provided the jitter is low enough \cite{hadfield2009}, this shows that a trade-off relationship exists independent of such strategies.  
At the same time, reducing the detection jitter incurs increased energetic costs of maintaining the detector at a "detection-ready" state, \emph{i.e.}~increased steady-state entropy production rate $\dot\Sigma_{ss}$ as can be seen from the colour bar plot of Fig.~\ref{fig:jitter_over_dark_count_rate_2}.

Intuitively, this two-fold trade-off between dark count rate, entropy production rate and jitter can be understood by noting that a sharp peak in detection current (facilitating low jitter) requires a highly populated metastable state at high energy. Driving the detector to this state and keeping it there can only be achieved with a strong driving bias, which requires negative virtual temperatures close to zero and thus incurs a high entropy production rate. At the same time, a highly excited metastable state is more susceptible to thermal excitations, resulting in an increased likelihood of dark counts. 

\begin{figure}
    \centering
    \includegraphics[width=1\linewidth]{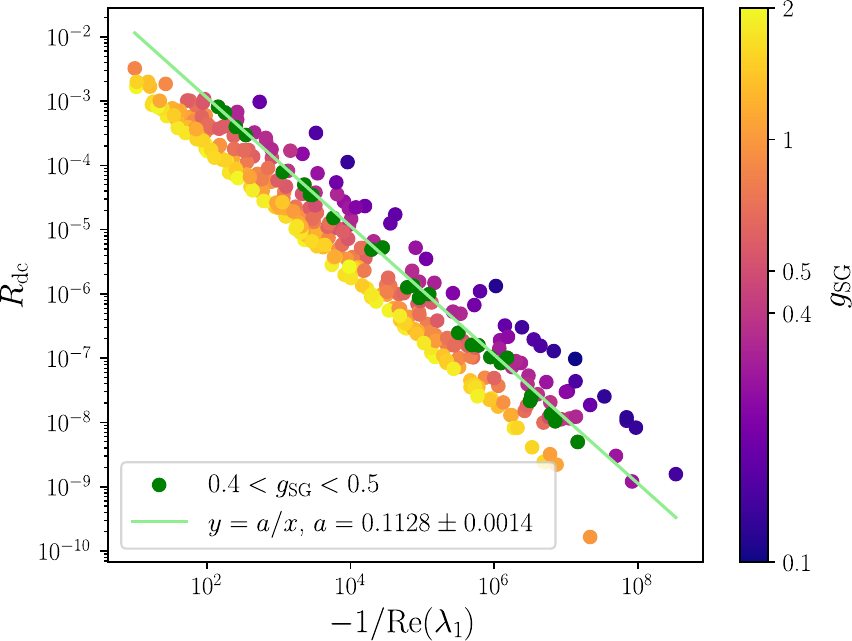}
    \caption{The dark count rate $R_\mathrm{dc}$ over $-1/\operatorname{Re}(\lambda_1)$, which is proportional to the dead time  (\emph{i.e.}~$-1/\operatorname{Re}(\lambda_1)\propto D$). The colour map shows different values of $g_\mathrm{SG}$. We consider data points close to maximal efficiency $0.72<\eta_\mathrm{D}<\eta_\mathrm{D}^\mathrm{max}\approx0,74$. The green line indicates a fit of $y=a/x$ where $x=1/\lambda_1$ and $y=R_\mathrm{dc}$. Decreasing the dead time comes at the cost of increased dark count rate. The parameter ranges for the scatter plots are the same as in Fig.~\ref{fig:Efficiency_entropy_production_rateD} except $g_\mathrm{SG}\in[0.1,2]$.} 
    \label{fig:dark_count_rate_over_inv_fist_gap}
\end{figure}

Finally, the speed at which information can be processed is intricately related to the rate at which it can be acquired. This rate is ultimately limited by the time it takes the detector to reset, \emph{i.e.}~the dead time (Eq.~\eqref{eq:epsilon_dead_time}).
We find that decreasing the dead time comes at the cost of increased dark count rate, shown numerically in Fig.~\ref{fig:dark_count_rate_over_inv_fist_gap}. This final trade-off thus takes the form
\begin{align}\label{eq:darkcount-deadtime}
    R_\mathrm{dc}\propto -\operatorname{Re}( \lambda_1) \sim \dfrac{1}{D}
\end{align}
Where the exact proportionality factor depends on specifics of the detector, \emph{e.g.} the coupling strength between system and gain medium, as depicted by the colour coding in Fig.~\ref{fig:dark_count_rate_over_inv_fist_gap}.

\begin{figure}[b]
    \centering
    \includegraphics[width=0.7\linewidth]{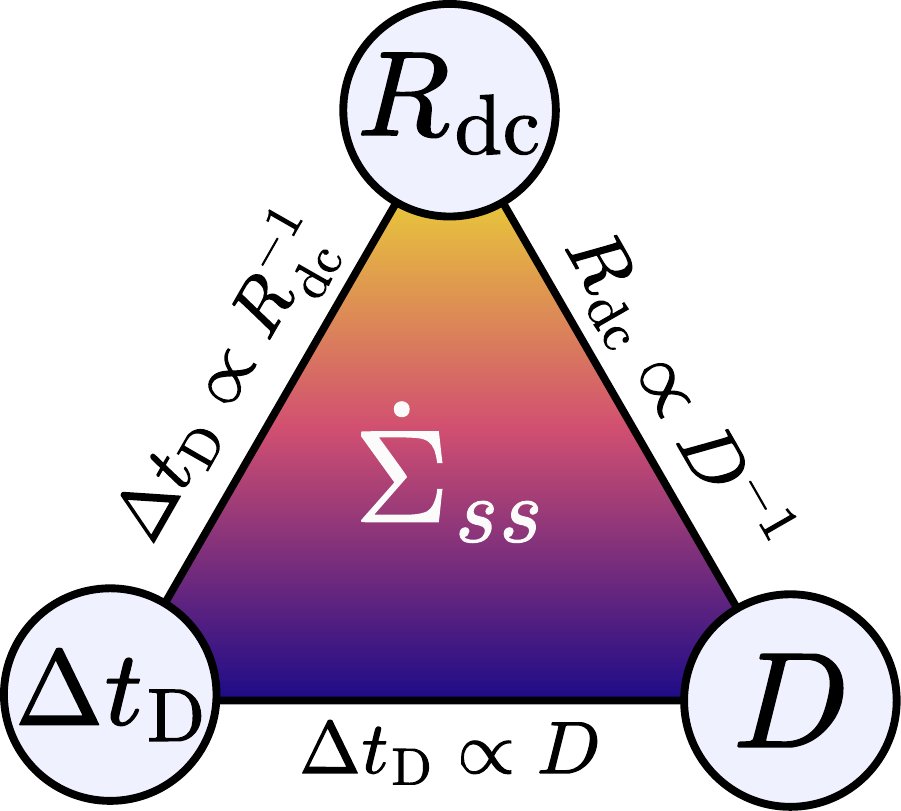}
    \caption{Illustration of the trade-off relations between dark count rate $R_\mathrm{dc}$, jitter $\Delta t_\mathrm{D}$ and dead time $D$. Increased $R_\mathrm{dc}$ comes with increased entropy production rate $\dot\Sigma_{ss}$ as indicated by the colour gradient. Because $R_\mathrm{dc}$ is inversely proportional to both $\Delta t_\mathrm{D}$ and $D$, proportionality between the latter two is implied (for numerical evidence for this see App.~\ref{appsec:jitter-deadtime}).}  
    \label{fig:triangle}
\end{figure}

Combining Eq.~\eqref{eq:darkcount-deadtime} and Eq.~\eqref{eq:jitter-darkcounts}, we find a three-fold relation between $R_\mathrm{dc}$, $\Delta t_\mathrm{D}$ and $D$, which is depicted in Fig.~\ref{fig:triangle}. The third relation, given by $\Delta t_\mathrm{D} \propto D$, is implied by the former two, where $g_\mathrm{SG}$ determines the proportionality constant. In App.~\ref{appsec:jitter-deadtime}, we show numerical evidence for this relation. Moving along the vertical axis of this trade-off triangle, \emph{i.e.} changing $R_\mathrm{dc}$ incurs a change in entropy production rate $\dot\Sigma_{ss}$, where increasing $R_\mathrm{dc}$ increases $\dot\Sigma_{ss}$. 

\section{Conclusion}
\label{sec:conclusions}

In this work, we have put forward a minimal autonomous model of a quantum particle detector. Our approach highlights that any such detector must be an open quantum system maintained in a metastable, non-equilibrium steady state (NESS) by a dissipative energy flux. As a result, good detectors irreversibly produce entropy. We have demonstrated that this entropy production constrains detection efficiency, in the sense that efficiency cannot be increased beyond a certain point without also increasing the total entropy produced per detection event. We have also shown that reducing the detection timing jitter requires increasing the steady-state entropy production rate, i.e.~the ``housekeeping heat'' needed to maintain the NESS~\cite{Seifert2012}. These results establish how the quality of the detection process is determined by thermodynamic irreversibility within the detector. 

While it was known that linear-response detectors must be out of equilibrium~\cite{clerk_introduction_2010}, our work is the first to explore the quantitative relation between entropy production and detector performance for particle detection: an inherently non-linear process. Our overall conclusions add to the consensus~\cite{guryanova2020,faist2015a, Chu2022, Sagawa2009,Granger2011,Jacobs2012, Latune2025, allahverdyan_quantum_2001, allahverdyan_curie-weiss_2003,schwarzhans2023a,engineer2024} that the acquisition of information comes at a thermodynamic cost. Here, however, our analysis uniquely emphasizes that the cost of measurement is fundamentally linked to the far-from-equilibrium nature of the detector. This cost can have significant consequences for quantum thermodynamics: for example, accurate measurements of work may themselves consume an amount of work far in excess of the measured work values~\cite{Debarba2019}. 

To identify the fundamental limits, our model assumes only limited complexity, i.e., with the minimal number of involved states and transitions. Real detectors are, of course, more complex and will exhibit many additional sources of parasitic dissipation that do not contribute to improving detector performance. 
However, given the recent insights that quantum clocks can circumvent entropic costs in accurate timing \cite{meier2025}, one may wonder whether more complex detector designs could exploit coherent many-body dynamics to achieve higher thermodynamic efficiency. We also note that here we focus on the thermodynamic cost of amplification once the particle has been captured by the detector, whereas the process of capturing an excitation (related to \ref{enum:physical_interaction}) may itself carry additional costs and inefficiencies.

Much work is still needed on the foundations of the measurement process and its energetic implications.
While the formalism of Lindblad master equations we use is fully compatible with an `all-unitary' picture of the measurement process, any stochastic collapse due to modifications of quantum mechanics~\cite{GRW,Oppenheim2023} would either leave our conclusions unchanged or add extra sources of entropy production~\cite{Cabello}. Unitarity of the process at the level of system and environment is also fully compatible with the measurement-equilibration hypothesis~\cite{schwarzhans2023a,engineer2024}, and an explicit connection between the two approaches can likely be made through the equivalence of quantum master equations with closed-system equilibration~\cite{ODonovan2025}. 

While we focus our thermodynamic analysis on the detection of a definite excitation, an attractive feature of our approach is that it can naturally deal with fields in a superposition of vacuum and a particle excitation, $a|0\rangle+b|1\rangle$. In that case, the proportion of trajectories with a detector click is equal to the efficiency multiplied by the squared coefficient of superposition $|b|^2$. This seamlessly encapsulates the Born rule, which follows from our model without any further fine-tuning or assumptions.

\raggedbottom

\section{Acknowledgments}
The authors thank Pharnam Bakhshinezhad, Florian Meier and Gabriel Landi for fruitful discussions. M. H. acknowledges support from ERC-2021-COG 101043705 ``Cocoquest''. This research was funded in whole or in part by the Austrian Science Fund (FWF) 10.55776/I6949, 10.55776/COE1 and the European Union – NextGenerationEU. M. T. M. is supported by a Royal Society University Research Fellowship. This project is co-funded by the European Union (Quantum Flagship project ASPECTS, Grant Agreement No.\ 101080167) and UK Research and Innovation (UKRI). Views and opinions expressed are however those of the authors only and do not necessarily reflect those of the European Union, Research Executive Agency or UKRI. Neither the European Union nor UKRI can be held responsible for them.

\bibliographystyle{apsrev4-1fixed_with_article_titles_full_names.bst}
\bibliography{bibfile.bib}
\onecolumngrid
\appendix

\renewcommand{\thesubsection}{\thesection.\arabic{subsection}}
\renewcommand{\thesubsubsection}{\thesubsection.\arabic{subsubsection}}
\section{Lindbladian}\label{appsec:Lindbladian}
The dynamics of the detector model is given via a GKLS master equation of the form
\begin{align}
    \mathcal{L}\rho=-i[H,\rho] + (\mathcal{D}_\mathrm{M_C} +\mathcal{D}_\mathrm{M_H}+\mathcal{D}_\mathrm{G}+\mathcal{D}_\mathrm{D})\rho,
\end{align}
where $H$ is the Hamiltonian of the system as given in the main text. The dissipators are given as follows:

The machine qubits are respectively interacting with the hot and the cold baths via
\begin{align}
    \mathcal{D}_\mathrm{M_{C/H}}=\sqrt{\Gamma_\mathrm{M}^+(T_{C/H})\,}\mathcal{D}[\sigma^+_\mathrm{M_{C/H}}]
     +\sqrt{\Gamma_\mathrm{M}^-(T_{C/H})\,}\mathcal{D}[\sigma^-_\mathrm{M_{C/H}}],
\end{align}
the gain medium interacts with the cold bath (a result of imperfect isolation in real-world settings) via
\begin{align}
    \mathcal{D}_\mathrm{G}=&\sqrt{\Gamma^-_{\mathrm{G}_{0-1}}(T_\mathrm{C})\,}\mathcal{D}[\ketbra{0}{1}_\mathrm{G}]+\sqrt{\Gamma^+_{\mathrm{G}_{0-1}}(T_\mathrm{C})\,}\mathcal{D}[\ketbra{1}{0}_\mathrm{G}]
    \\&
    +\sqrt{\Gamma^-_{\mathrm{G}_{1-2}}(T_\mathrm{C})\,}\mathcal{D}[\ketbra{1}{2}_\mathrm{G}]+\sqrt{\Gamma^+_{\mathrm{G}_{1-2}}(T_\mathrm{C})\,}\mathcal{D}[\ketbra{2}{1}_\mathrm{G}],
\end{align}
and the gain medium interacts with the detection channel via
\begin{align}
    \mathcal{D}_\mathrm{D}=\sqrt{\Gamma^+_\mathrm{D}(T_\mathrm{C})\,} \mathcal{D}[\ketbra{2}{0}_\mathrm{G}]+\sqrt{\Gamma^-_\mathrm{D}(T_\mathrm{C})\,} \mathcal{D}[\ketbra{0}{2}_\mathrm{G}].
\end{align}
 Each dissipator is given in terms of the jump $L$ operators as $\mathcal{D}[L]\rho=L \rho L^{\dagger}-\frac{1}{2}\left\{L^{\dagger} L, \rho\right\}$.
The corresponding temperature-dependent coupling rates are given by 
\begin{align}
    &\Gamma_\mathrm{M}^+(T_{C/H})=\gamma_\mathrm{M} \frac{e^{-E_{C/H}/T_{C/H}}}{\mathcal{Z}_{C/H}},& \Gamma^-_\mathrm{M}(T_{C/H})= \frac{\gamma_\mathrm{M}}{\mathcal{Z}_{C/H}}
    \\& 
    \Gamma^+_{\mathrm{G}_{0-1}}(T_\mathrm{C})=\gamma_\mathrm{G}(\dfrac{e^{-E_\mathrm{G}/T_\mathrm{C}}}{\mathcal{Z}_{\mathrm{G}_{0-1}}}), &  \Gamma^-_{\mathrm{G}_{0-1}}(T_\mathrm{C})=\gamma_\mathrm{G}(\dfrac{1}{\mathcal{Z}_{\mathrm{G}_{0-1}}})    
    \\& 
    \Gamma^+_{\mathrm{G}_{1-2}}(T_\mathrm{C})=\gamma_\mathrm{G}(\dfrac{e^{-E_\mathrm{S}/T_\mathrm{C}}}{\mathcal{Z}_{\mathrm{G}_{1-2}}}), &  \Gamma^-_{\mathrm{G}_{1-2}}(T_\mathrm{C})=\gamma_\mathrm{G}(\dfrac{1}{\mathcal{Z}_{\mathrm{G}_{1-2}}})
    \\& 
    \Gamma^+_{\mathrm{G}_\mathrm{D}}(T_\mathrm{C})=\gamma_\mathrm{D}(\dfrac{e^{-(E_\mathrm{S}+E_\mathrm{G})/T_\mathrm{C}}}{\mathcal{Z}_{\mathrm{G}_\mathrm{D}}}), &  \Gamma^-_{\mathrm{G}_\mathrm{D}}(T_\mathrm{C})=\gamma_\mathrm{D}(\dfrac{1}{\mathcal{Z}_{\mathrm{G}_\mathrm{D}}})
\end{align}
with the partition functions corresponding to the respective energy gaps $E_x$ at bath temperature $T_x$ given as ${\mathcal{Z}_x=1+e^{-E_x/T_x}}$.
\section{Figures of Merit}\label{appsec:figuresofmerit}
As outlined in the introduction, an ideal detector should fulfill desiderata \ref{enum:dark_counts}, \ref{enum:jitter}, \ref{enum:efficiency}, and \ref{enum:deadtime}. These properties determine the figures of merit typically quantifying the performance of detectors (see \emph{e.g.} for photon detection \cite{hadfield2009}): \emph{dark count rate}, \emph{timing jitter} and \emph{detection efficiency}.

As a prerequisite, we would like to remind the reader of the following notation:
In order to simplify notation, we will transform all operators into vectors in Liouville space, stacking their columns in the following way
\begin{align}
    \operatorname{vec}\left(\begin{array}{ll}a & b \\ c & d\end{array}\right)=\left(\begin{array}{l}a \\ c \\ b \\ d\end{array}\right)
\end{align}
using the notation
\begin{align}
    |\rho\rangle\!\rangle=\operatorname{vec}(\rho),
\end{align}
and all superoperators into operators using the convention 
\begin{align}
    \operatorname{vec}(A \rho C)=\left(C^{\mathrm{T}} \otimes A\right) \operatorname{vec}(\rho)
\end{align}
Using this, the current through some channel $x$ can be expressed in Liouville space via the superoperator for current $\mathcal{J}$ as 
\begin{align}
    J_x(t)=\langle\!\langle \mathds{1}|\mathcal{J}_x|\rho(t)\rangle\!\rangle,
\end{align}
where $\langle\!\langle \mathds{1}|$ is the vectorised identity operator. The time-dependent state can be written as
\begin{align}
    |\rho(t)\rangle\!\rangle =|\rho^{ss}\rangle\!\rangle + \sum_{i>0}e^{\lambda_i t}|x_i\rangle\!\rangle \langle\!\langle y_i|\rho_0\rangle\!\rangle,
\end{align}
where $|x_i\rangle\!\rangle$ and $ \langle\!\langle y_i|$ are the right and left eigenvectors of the Liouvillian corresponding to eigenvalue $\lambda_i$ (with $\operatorname{Re}(\lambda_i)< 0$), fulfilling $\langle\!\langle y_i|x_i\rangle\!\rangle=\delta_{ij}$ (note that these are not complex conjugations of each other as the Liouvillian is non-hermitian, so $|x_i\rangle\!\rangle^\dagger\neq\langle\!\langle y_i|$), and $|\rho_0\rangle\!\rangle$ is the initial state. 

The Drazin pseudo-inverse of the Liouvillian is defined as 
\begin{align}\label{eq:drazinverse}
   \mathcal{L}^{+}:=\sum_{i > 0} \frac{1}{\lambda_i}|x_i\rangle\!\rangle \langle\!\langle y_i|~.
\end{align}
We will denote the transient ``excess current" through channel $z$ as $\tilde{J}_z(t)$ and rewrite in the Liouville space as
\begin{align}\label{eq:excesscurrent}
    \tilde{J}_z(t)=J_z(t)-J_z^{ss}=\langle\!\langle \mathds{1}|\mathcal{J}_z|\rho(t)\rangle\!\rangle-\langle\!\langle \mathds{1}|\mathcal{J}_z|\rho^{ss}\rangle\!\rangle=\langle\!\langle \mathds{1}|\mathcal{J}_z\sum_{i>0}e^{\lambda_i t}|x_i\rangle\!\rangle \langle\!\langle y_i|\rho_0\rangle\!\rangle
\end{align}

\subsection{Detection efficiency}

The \emph{detection efficiency} $\eta_\mathrm{D}$ is defined as the ratio of counts recorded to the total number of counts that could have been detected \cite{hadfield2009}
\begin{align}\label{eq:efficiency_rates}
    \eta_\mathrm{D}=\dfrac{\#\text{detection events}}{\#\text{incident events}}=\dfrac{N_\mathrm{D}}{N_\mathrm{I}}
\end{align}
The number of counts detected $N_\mathrm{D}$ can be equivalently expressed as the number of total incident events $N_\mathrm{I}$ times $P_\text{Detection}$, the probability per incident event to have a detection event, \emph{i.e.}
\begin{align}\label{eq:efficiency}
    N_\mathrm{D}=  N_\mathrm{I} P_\text{Detection}= N_\mathrm{I} \int_0^\infty dt(J_\mathrm{D}(t)-J^{ss}_\mathrm{D}) 
\end{align}
Where the second equality comes from each detection event being i.i.d. in our model (based on the assumption that there are very few quantum particles around and the detector always resets perfectly after each detection event). As a consequence, the detection efficiency for a single incident event is given as
\begin{align}
   \eta_\mathrm{D}=P_\text{Detection}=\int_0^\infty dt(J_\mathrm{D}(t)-J^{ss}_\mathrm{D}) =\int_0^\infty dt \tilde J_\mathrm{D}(t)
\end{align}
where $J_\mathrm{D}(t)$ is the detection current, $J^\mathrm{ss}_\mathrm{D}$ is the steady state detection current (and thus the current flowing when there is no particle). To understand this one can think of the ``net detection current" $J_\mathrm{D}(t)-J^{ss}_\mathrm{D}$ as a rate of detection whose integral gives exactly the average number of jumps in the detection channel that exceed the number of jumps that would be present in the steady state. Using eq.~\eqref{eq:excesscurrent} and eq.~\eqref{eq:drazinverse} the efficiency can be written as
\begin{align}
    \eta_\mathrm{D}&=\int_0^\infty \tilde{J}_\mathrm{D}(t)dt=\int_0^\infty dt\langle\!\langle \mathds{1}|\mathcal{J_\mathrm{D}}\sum_{i>0}e^{\lambda_i t}|x_i\rangle\!\rangle \langle\!\langle y_i|\rho_0\rangle\!\rangle= \langle\!\langle \mathds{1}|\mathcal{J_\mathrm{D}}\sum_{i>0}\int_0^\infty dt e^{\lambda_i t} |x_i\rangle\!\rangle \langle\!\langle y_i|\rho_0\rangle\!\rangle
    \\&= - \langle\!\langle \mathds{1}|\mathcal{J_\mathrm{D}}\sum_{i>0} \dfrac{1}{\lambda_i} |x_i\rangle\!\rangle \langle\!\langle y_i|\rho_0\rangle\!\rangle= - \langle\!\langle \mathds{1}|\mathcal{J_\mathrm{D}}\mathcal{L}^+|\rho_0\rangle\!\rangle, \label{eq:eff_deriv_2}
\end{align}
where we used $\int_0^\infty dt\; t^n e^{\lambda_i  t}=(- \lambda_i)^{1-n} n!$ for $n\in \mathds{N}\, \backslash \, \{ 0\}$, and $\langle\!\langle y_i|x_j\rangle\!\rangle=\delta_{ij}$. Thus, the detection efficiency is given by
\begin{align}
    \eta_\mathrm{D}=-\langle\!\langle \mathds{1}|\mathcal{J_\mathrm{D}}\mathcal{L}^+|\rho_0\rangle\!\rangle
\end{align}
and requires only the diagonalisation of the Liouvillian.

\subsection{Dark count rate}

As the dynamics do not feature any process by which the system can be excited stochastically, \emph{i.e.}~no external excitations with energy $E_S$ can occur, once the detector reaches its non-equilibrium steady state, we assume that the remaining current in the detection channel can not be the result of actual detection events but has to come from thermal fluctuations.  The rate at which these ``false positive" detection counts occur is called \emph{dark count rate}, and is given as \begin{align}
    R_\mathrm{dc}=J^{ss}_\mathrm{D}=\langle\!\langle \mathds{1}|\mathcal{J_\mathrm{D}}|\rho^{ss}\rangle\!\rangle
\end{align}

\subsection{Detection jitter}
Detection of quantum particles amounts to answering the question: Was there a quantum particle present at the \emph{location of the detector} at a certain \emph{point in time}? Detection thus corresponds to the localisation of an event in space and time. Ideally, a detector clicks exactly at the moment in time when the particle hits it. However, in practice, the detection time fluctuates. These fluctuations can be quantified via the \emph{detection jitter}, which is typically benchmarked for real-world detectors using the full width at half maximum (FWHM) of the detector's instrument response function \cite{hadfield2009}. Here we will refer to the jitter as the variance of the excess current $\tilde{J}_\mathrm{D}(t)$. 
 \begin{align}
     \Delta t_\mathrm{D} &=\operatorname{var}(t)_{p_{\tilde{J}_\mathrm{D}}}=\int_0^\infty dt\,t^2  \frac{\tilde J_\mathrm{D}(t)}{\eta_\mathrm{D}} -\left[\int_0^\infty dt\,t \frac{\tilde J_\mathrm{D}(t)}{\eta_\mathrm{D}}\right]^2 
     \\& =\langle\!\langle \mathds{1}|\dfrac{\mathcal{J_\mathrm{D}}}{\eta_\mathrm{D}}\sum_{i>0}\int_0^\infty dt (e^{\lambda_i t}t^2)|x_i\rangle\!\rangle \langle\!\langle y_i|\rho_0\rangle\!\rangle-\left[\langle\!\langle \mathds{1}|\dfrac{\mathcal{J_\mathrm{D}}}{\eta_\mathrm{D}}\sum_{i>0}\int_0^\infty dt (e^{\lambda_i t}t)|x_i\rangle\!\rangle \langle\!\langle y_i|\rho_0\rangle\!\rangle\right]^2
     \\&= -2\langle\!\langle \mathds{1}|\dfrac{\mathcal{J_\mathrm{D}}}{\eta_\mathrm{D}}(\mathcal{L}^+)^3|\rho_0\rangle\!\rangle-\left[\langle\!\langle \mathds{1}|\dfrac{\mathcal{J_\mathrm{D}}}{\eta_\mathrm{D}}(\mathcal{L}^+)^2|\rho_0\rangle\!\rangle\right]^2.
 \end{align}
 where we used the same integration as in eq.~\eqref{eq:eff_deriv_2}. The factor $\tfrac{1}{\eta_\mathrm{D}}$ is needed since the variance requires a correctly normalised distribution to be calculated. Since $\tilde J_\mathrm{D}(t)$ is not normalised, we normalise it: ${\tilde J_\mathrm{D}(t)/\int_0^{\infty}dt \tilde J_\mathrm{D}(t) = \tilde J_\mathrm{D}(t)/\eta_\mathrm{D}}$.
This leaves us with the detection jitter
\begin{align}
    \Delta t_\mathrm{D} = -2\langle\!\langle \mathds{1}|\dfrac{\mathcal{J_\mathrm{D}}}{\eta_\mathrm{D}}(\mathcal{L}^+)^3|\rho_0\rangle\!\rangle-\left(\langle\!\langle \mathds{1}|\dfrac{\mathcal{J_\mathrm{D}}}{\eta_\mathrm{D}}(\mathcal{L}^+)^2|\rho_0\rangle\!\rangle\right)^2,
\end{align}
which again depends solely on properties of the Liouvillian. 

\subsection{Dead time}
For any task that requires a high rate of measurements, it is important to quantify the minimal time at which measurements can be reliably performed, \emph{i.e.}~the \emph{dead time} of the detector $D$. Once the detection event has happened, the detector returns to its equilibrium state, ready for the next detection event. The time this takes is generally related to the first gap of the Liouvillian. One can write the time-evolved state in vectorised notation as
\begin{align}
    |\rho(t)\rangle\!\rangle=\sum_{i>0} e^{\lambda_i t}|x_i\rangle\!\rangle\!\langle\!\langle y_i|\rho_0\rangle\!\rangle+|\rho^{ss}\rangle\!\rangle
\end{align}
since $\operatorname{Re}(\lambda_{i+1})\leq \operatorname{Re}(\lambda_i)$ and $\operatorname{Re}(\lambda_i)\leq 0 \forall i$, we can ignore all eigenvalues (and corresponding modes) larger than the second one (the first is zero and corresponds to the steady state) as they all decay faster, and define the dead time via the slowest decaying mode as
\begin{align}
    D\sim -\dfrac{1}{\operatorname{Re}(\lambda_1)}.
\end{align}
\subsection{Total entropy production}\label{appsec:entropy}
In a similar fashion to the efficiency, we can calculate the transient excess entropy production $\Sigma_\mathrm{trans}$. For weak interactions within the system compared to the interaction with the baths (to have well-defined energies transported in each jump), it is given by
\begin{align}
    \Sigma_\mathrm{trans}=\dfrac{1}{T_\mathrm{C}}\int_0^\infty dt \left( E_\mathrm{C} \tilde J_\mathrm{M_C}(t) + E_\mathrm{G} \tilde J_\mathrm{G_{1-0}}(t)+E_\mathrm{S}\tilde J_\mathrm{G_{2-1}}(t)\right)=\dfrac{1}{T_\mathrm{C}}\langle\!\langle \mathds{1}|(E_\mathrm{C} \mathcal{J}_\mathrm{M_C}+ E_\mathrm{G}\mathcal{J}_\mathrm{G_{1-0}}+E_\mathrm{S} \mathcal{J}_\mathrm{G_{2-1}})\mathcal{L}^+ |\rho_0\rangle\!\rangle
\end{align}

Keeping the detector ready for detection requires it to be constantly driven out of equilibrium, which comes at its own cost, which we call the steady-state entropy production rate $\dot\Sigma_{ss}$ and is given by
\begin{align}
    \dot\Sigma_{ss} =\dfrac{1}{T_\mathrm{C}}\langle\!\langle \mathds{1}|(E_\mathrm{C} \mathcal{J}_\mathrm{M_C}+ E_\mathrm{G}\mathcal{J}_\mathrm{G_{1-0}}+E_\mathrm{S} \mathcal{J}_\mathrm{G_{2-1}})|\rho^{ss}\rangle\!\rangle
\end{align}

The total entropy production per detection event $\Sigma_\mathrm{tot}$ can be calculated by taking the entropy produced in the timescale over which one detection event takes place, given by $D \dot\Sigma_{ss}\sim-\dot\Sigma_{ss}/\operatorname{Re}(\lambda_1) $, and adding the transient excess entropy production, \emph{i.e.}
\begin{align}
  \Sigma_\mathrm{tot}\sim -\frac{\dot\Sigma_{ss}}{\operatorname{Re}(\lambda_1)} + \Sigma_\mathrm{trans}
\end{align}

\section{Efficiency bound hypothesis}\label{appsec:Efficiency_bound}
In this section, we present an argument for the validity of the upper bound in eq.~\eqref{eq:max_efficiency}. 
First note that the efficiency, as defined in eq.~\eqref{eq:efficiency_rates} can be written as 
\begin{align}
    \eta_\mathrm{D}=\dfrac{N_\mathrm{D}}{N_\mathrm{D}+N_\mathrm{Lost}},
\end{align}
where $N_\mathrm{Lost}$ counts all the particles that were not detected, such that $N_\mathrm{I}=N_\mathrm{D}+N_\mathrm{Lost}$. The only way an excitation can leave the subspace spanned by $\ket{2}_\mathrm{G}\otimes \ket{0}_S$ and $\ket{1}_\mathrm{G}\otimes \ket{1}_S$ is via the detection channel $\mathcal{D}_\mathrm{D}$ inducing the jump $\ket{2}_\mathrm{G}\rightarrow \ket{0}_\mathrm{G}$ or via the dissipative coupling to the cold bath $\ket{2}_\mathrm{G}\rightarrow \ket{1}_\mathrm{G}$. Thus, all counts in $N_\mathrm{Lost}$ must be accounted for in the current between the $\ket{1}-\ket{2}$ gap in the gain medium and the bath, given by
\begin{align}
    J_{\mathrm{G}_{1-2}}(t)=\operatorname{Tr}[(\Gamma_{\mathrm{G}_{1-2}}^+\ketbra{1}{1}_\mathrm{G} -\Gamma_{\mathrm{G}_{1-2}}^-\ketbra{2}{2}_\mathrm{G})\rho(t)].
\end{align}
minus the corresponding steady state current $J_{\mathrm{G}_{1-2}}^{ss}$, \emph{i.e} $N_\mathrm{Lost}=\int_0^\infty (J_{\mathrm{G}_{1-2}}(t)-J_{\mathrm{G}_{1-2}}^{ss})dt$. The total efficiency can thus be written as 
\begin{align}
    \eta_\mathrm{D}=\dfrac{\int_0^\infty dt(J_\mathrm{D}(t)-J_\mathrm{D}^{ss})}{\int_0^\infty dt(J_\mathrm{D}(t)-J_\mathrm{D}^{ss})+\int_0^\infty dt(J_{\mathrm{G}_{1-2}}(t)-J_{\mathrm{G}_{1-2}}^{ss})}
\end{align}
We will write these two integrals in simplified notation as
\begin{align}
    &N_\mathrm{D}=\int_0^\infty dt (J_\mathrm{D}(t)-J_\mathrm{D}^{ss})=I_2\Gamma_{D}^- - I_0\Gamma_{D}^+ 
    \\&
    N_\mathrm{Lost}=\int_0^\infty  dt (J_{\mathrm{G}_{1-2}}(t)-J_{\mathrm{G}_{1-2}}^{ss})=I_2\Gamma_{\mathrm{G}_{1-2}}^- - I_1\Gamma_{\mathrm{G}_{1-2}}^+
\end{align}
where $I_2\Gamma_{\mathrm{D}/{\mathrm{G}_{1-2}}}^-=\int_0^\infty dt(P_2(t)-P_2^{ss})\Gamma_{\mathrm{D}/{\mathrm{G}_{1-2}}}^- dt $ and $I_{0/1}\Gamma_{D/{\mathrm{G}_{1-2}}}^+=\int_0^\infty dt(P_{0/1}(t)-P_{0/1}^{ss})\Gamma_{D/{\mathrm{G}_{1-2}}}^+ dt $, and $P_i(t)$ ($P_i^{ss}$) correspond to $\operatorname{Tr}(\rho(t) \ketbra{i}{i}_\mathrm{G})$ ($\operatorname{Tr}(\rho^{ss} \ketbra{i}{i}_\mathrm{G})$ ).
Thus, the efficiency can be expressed as 
\begin{align}
    \dfrac{1}{1+ \dfrac{I_2\Gamma_{{\mathrm{G}_{1-2}}}^- - I_1\Gamma_{{\mathrm{G}_{1-2}}}^+}{I_2\Gamma_{D}^- - I_0\Gamma_{D}^+}}
\end{align}
In the zero temperature case, i.e. when $T_\mathrm{C}\rightarrow0$, using the coupling rates from App.~\ref{appsec:Lindbladian}, we have 
\begin{align}
    \eta_\mathrm{D} \rightarrow \dfrac{1}{1+ \frac{I_2\Gamma_{{\mathrm{G}_{1-2}}}^- }{I_2\Gamma_{D}^- }}  =  \dfrac{1}{1+ \frac{\gamma_{{\mathrm{G}}} }{\gamma_{D} }}
\end{align}
which is equivalent to eq.~\eqref{eq:max_efficiency}. In the following, we show that this is, in fact, an upper bound for the efficiency. To show this, we must show that
\begin{align}
    \dfrac{1}{1+ \dfrac{I_2\Gamma_{{\mathrm{G}_{1-2}}}^- - I_1\Gamma_{{\mathrm{G}_{1-2}}}^+}{I_2\Gamma_{D}^- - I_0\Gamma_{D}^+}} \leq \dfrac{1}{1+ \frac{\gamma_{\mathrm{G}} }{\gamma_{D}}}
\end{align}
or equivalently that 
\begin{align}
    I_1\dfrac{\Gamma_{{\mathrm{G}_{1-2}}}^+}{\gamma_{{\mathrm{G}}}}\leq I_0\dfrac{\Gamma_{\mathrm{D}}^+}{\gamma_{\mathrm{D}}}
\end{align}
substituting in the coupling rates, and bringing all of them to the right, and noting, that $\tfrac{\gamma_\mathrm{G}\Gamma_{\mathrm{D}}^+}{\Gamma_{{\mathrm{G}_{1-2}}}^+\gamma_\mathrm{D}} =e^{-E_\mathrm{G}/T_\mathrm{C}}\leq 1$ we get the condition that
\begin{align}\label{eq:I1<I0}
    I_1 \leq I_0 e^{-E_\mathrm{G}/T_\mathrm{C}}\leq I_0
\end{align}
To see why we believe that this inequality should hold (apart from numerical evidence), consider the initial state of the system:
\begin{align}
    \rho_0=\rho_\mathrm{MG}^{ss}\otimes\ketbra{1}{1}_\mathrm{S},
\end{align}
where $\rho_\mathrm{MG}^{ss}$ is the steady state of the machine-gain medium subsystem, and the system qubit is prepared in its excited state. This choice of initial state creates a strong excitation imbalance relative to the global steady-state, with all population initially in level $\ket{1}_S$, and none in $\ket{0}_S$.

At early times, this imbalance drives population transfer from $\ket{1}_\mathrm{G}$ to $\ket{2}_\mathrm{G}$ due to the coherent coupling with the gain medium. As a result, the population $P_2(t)$ rises rapidly above its steady-state value, while $P_1(t)$ drops sharply below its steady-state value. Since $\ket{2}_\mathrm{G}$ decays into $\ket{0}_\mathrm{G}$ via the detection channel, the surplus in $P_2$ directly leads to a transient surplus in $P_0(t)$, i.e., $P_0(t) > P_0^{ss}$ for an extended period. At the same time, the transient depletion of $P_1(t)$ is counteracted by the thermal two-qubit engine, which is pumping the excess population in $P_0(t)$ that occurred due to decay via the detection channel $\ket{2}\rightarrow\ket{0}$ into $\ket{1}$. This feedback process counteracts the depletion in $P_1$ but occurs over longer timescales due to this conditional nature. As a result, while $P_1(t)$ initially drops below $P_1^{ss}$ and eventually recovers, the time-integrated deviation $(P_1(t) - P_1^{ss})$ remains negative for a larger portion of the evolution compared to $(P_0(t) - P_0^{ss})$, which is predominantly positive. Therefore, we hypothesise that eq.~\eqref{eq:I1<I0} is satisfied.

The equations of motion for the populations of the gain medium are given by
\begin{align}
& \dot{P}_2(t)=2 g_\mathrm{SG} \operatorname{Im}\left(\rho_{11,20}^\mathrm{SG}\right)-\Gamma_\mathrm{D}^{-} P_2+\Gamma_\mathrm{D}^{+} P_0-\Gamma_{\mathrm{G}_{1-2}}^{-} P_2+\Gamma_{\mathrm{G}_{1-2}}^{+} P_1 \\
& \dot{P}_1(t)=2\left(g_\mathrm{MG} \operatorname{Im}\left(\rho_{010,101}^\mathrm{MG}\right)-g_\mathrm{SG} \operatorname{Im}\left(\rho_{11,20}^\mathrm{SG}\right)\right)+\Gamma_{\mathrm{G}_{1-2}}^{-} P_2-\Gamma_{\mathrm{G}_{1-2}}^{+} P_1-\Gamma_{\mathrm{G}_{0-1}}^{-} P_1+\Gamma_{\mathrm{G}_{0-1}}^{+} P_0 \\
& \dot{P}_0(t)=-2 g_\mathrm{MG} \operatorname{Im}\left(\rho_{010,101}^\mathrm{MG}\right)+\Gamma_\mathrm{D}^{-} P_2-\Gamma_\mathrm{D}^{+} P_0+\Gamma_{\mathrm{G}_{0-1}}^{-} P_1-\Gamma_{\mathrm{G}_{0-1}}^{+} P_0\\
& P_0(t)+P_1(t)+P_2(t)=1
\end{align}


\section{Detection jitter-dead time relation}\label{appsec:jitter-deadtime}
\begin{figure}[htbp!]
    \centering
    \includegraphics[width=0.5\linewidth]{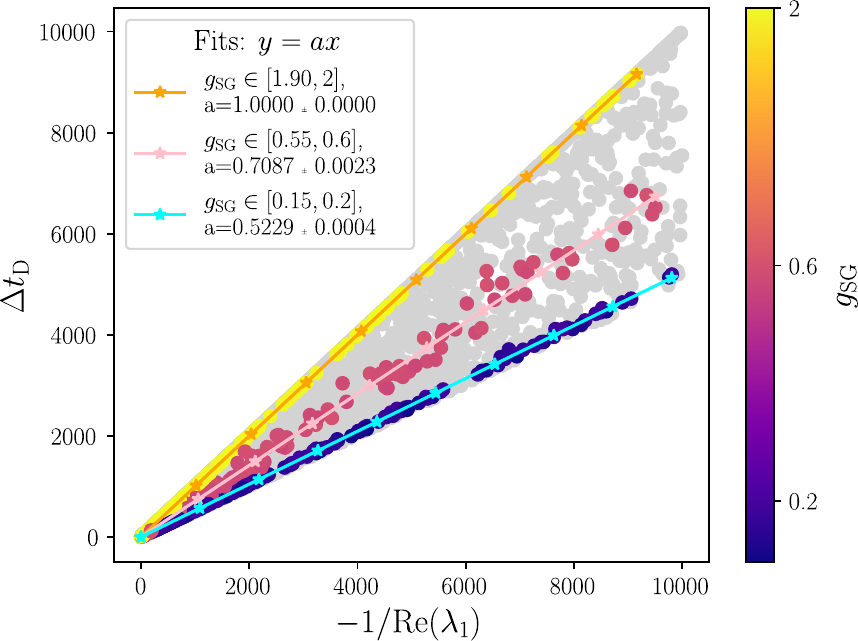}
    \caption{Jitter over the inverse first gap of the Liouvillian $1/\lambda_1$. The grey points display the full dataset where we restrict to cases where $\eta_\mathrm{D}> 0.4$. The coloured points indicate the data points selected for the fit by restricting $g_\mathrm{SG}$. The lines with stars indicate the linear fits.}
    \label{fig:jitter_over_inv_first_gap}
\end{figure}

This appendix aims to highlight the relationship between dead time and detection jitter. Intuitively, it is clear that detection jitter must always be smaller than the timescale at which equilibration happens (which corresponds to the dead time, see eq.~\eqref{eq:epsilon_dead_time} and the preceding paragraph), since the peak of the detection current must be within this time. Indeed, we find that jitter and inverse first gap $1/\lambda_1$ follow a linear scaling relation in this model, which is shown in the scatter plot in Fig.~\ref{fig:jitter_over_inv_first_gap}. The slope of this linear relationship increases with the system-gain medium interaction strength $g_{sl}$, approaching a maximal value of $1$. Using the fact that the dead time is bounded by the equilibration time, which is of the order of $1/\lambda_1$, our numerical evidence points us to the relation
\begin{align}
    \Delta t_\mathrm{D}\propto -\dfrac{1}{\operatorname{Re}(\lambda_1)}= D.
\end{align}

\section{Entropic cost of increasing the gain}\label{appsec:gain}
\begin{figure}
    \centering
    \begin{minipage}{0.47\linewidth}
        \centering
        \includegraphics[width=\linewidth]{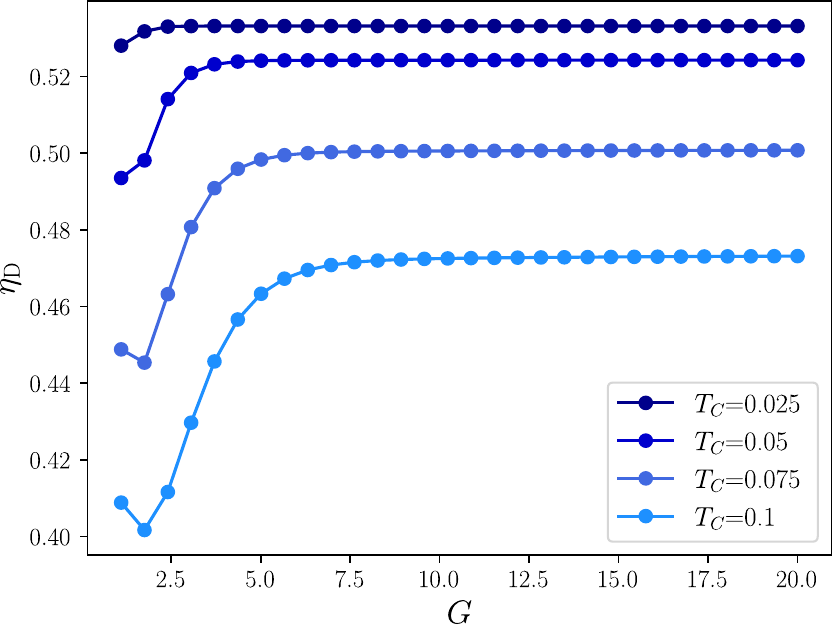}        \label{fig:efficiency_over_emax}
    \end{minipage}
    \begin{minipage}{0.47\linewidth}
        \centering
        \includegraphics[width=\linewidth]{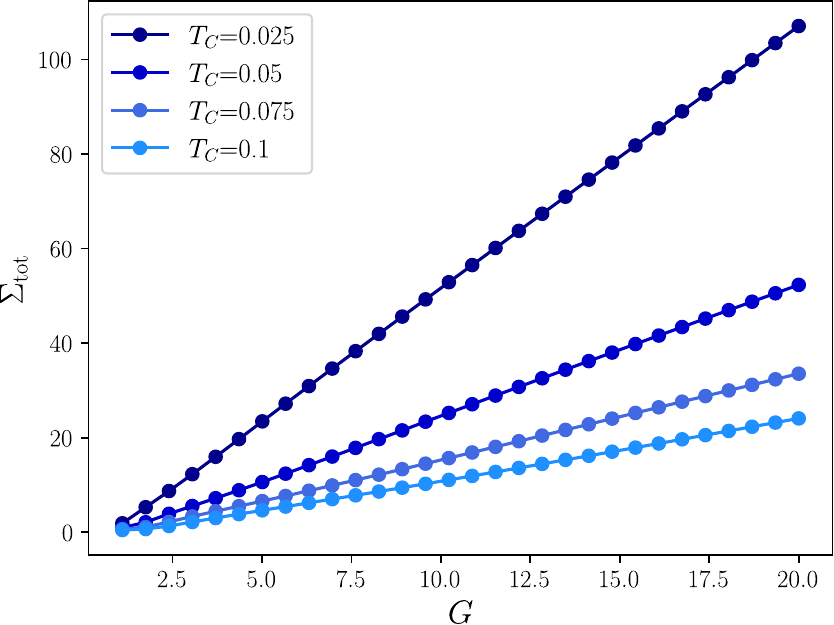}
        \label{fig:entropy_total_over_emax}
    \end{minipage}
\caption{Left panel: Detection efficiency  $\eta_\mathrm{D}$ over gain $G$ for different values of $T_C$. The initial lower efficiency can be explained by increased thermal fluctuations for a smaller gap size. As the first gap increases, these fluctuations are suppressed. For low enough $T_C$, efficiency and gain are independent of each other, \emph{i.e.}, increasing the gain does not affect the efficiency. Right panel: Total entropy production $\Sigma_\mathrm{tot}$ over energy gain $G$ for varying cold bath temperatures $T_C$. Increasing the gain comes with a linear increase in total entropy production. Here we fixed $E_\mathrm{S}$ and increased $E_\mathrm{G}$. Both plots were generated for one detector realisation with parameters $g_\mathrm{SG}=1$, $g_\mathrm{MG}=1$, $\gamma_\mathrm{D}=0.8$, $\gamma_\mathrm{G}=0.7$, $\gamma_\mathrm{M}=1$, $T_V=-3$, $f_{E_\mathrm{C}}=0.2$ ($E_\mathrm{C}=f_{E_\mathrm{C}}E_\mathrm{L}$).}
    \label{fig:Gain}
\end{figure}

In this section, we discuss the scaling of entropy costs of increased energy gains. In Fig.~\ref{fig:Gain} we show numerical evidence for a linear scaling of the total entropy production $\Sigma_\mathrm{tot}=\Sigma_\mathrm{trans} + \dot\Sigma_{ss}/\lambda_1$  with the gain $G=\tfrac{E_\mathrm{G}}{E_\mathrm{S}}$, \emph{i.e.}~the amount of energy amplification in this model, while leaving the efficiency unchanged for large enough gain values.

\end{document}